\documentclass{ws-procs975x65}

\def\beq{\begin{equation}}
\def\eeq{\end{equation}}

\newcommand{\mb}[1]{\mathbf{#1}}

\begin{document}

\title{Black hole thermodynamics in finite time}
\author{Christine Gruber$^*$}

\address{Institute of Physics, Carl v. Ossietzky University of Oldenburg,\\
26129 Oldenburg, Germany\\
$^*$E-mail: christine.gruber@uni-oldenburg.de}

\begin{abstract}
Finite-time thermodynamics provides the means to revisit ideal thermodynamic equilibrium processes in
the light of reality and investigate the energetic "price of haste", i.e. the consequences of carrying 
out a process in finite time, when perfect equilibrium cannot be awaited due to economic
reasons or the nature of the process. Employing the formalism of geometric thermodynamics, a lower bound
on the energy dissipated during a process is derived from the thermodynamic length of that process.
The notion of length is hereby defined via a metric structure on the space of equilibrium thermodynamics,
spanned by a set of thermodynamic variables describing the system. 
Since the aim of finite-time thermodynamics is to obtain realistic limitations on idealized scenarios, it
is a useful tool to reassess the efficiency of thermodynamic processes. We examine its implications for 
black hole thermodynamics, in particular scenarios inspired by the Penrose process, a thought experiment by 
which work can be extracted from a rotating black hole. We consider a Kerr black hole which, by some mechanism, 
is losing mass and angular
momentum. Thermodynamically speaking, such a process is described in the equilibrium phase space of the
black hole, but in reality, it is neither reversible nor infinitely slow. We thus calculate the dissipated 
energy due to non-ideal finite-time effects. 
\end{abstract}


\bodymatter

\section{Introduction}
Even though experimentally black holes have so far remained evasive and mysterious astrophysical objects, black hole 
thermodynamics is an active and innovative field of research. In the recent years black holes have been approached 
from an increasingly engineering point of view. Investigations in this direction include studies of black hole stability 
with respect to thermodynamic fluctuations or phase transitions, and proposals of heat engines involving black holes. 

One of the first ideas to utilize black holes as engines to gain energy from was put forward in the 1970s, suggesting the 
extraction of angular momentum from a rotating black hole \cite{penrose1971extraction}. Since then, many different mechanisms 
have been contemplated in this context, usually employing a specific mechanical process (for a recent review, see 
Ref.\,\refcite{2015Brav}). In this work, we suggest an approach based on thermodynamics, retaining maximum generality which 
does not require fixing a particular method of energy extraction. 

We will furthermore include the effects of finite time into our analysis, i.e. consider realistic thermodynamic processes 
\cite{andresen2011current} as opposed to idealised reversible ones. In 
reversible thermodynamics it is assumed implicitly that processes happen ideally and are represented by a sequence of perfect 
equilibrium states along the path - which requires an infinite amount of time. In reality however, processes are carried out on 
finite time scales, and thus at each local equilibrium state along the sequence a small amout of energy is lost, i.e. a small 
amount of entropy $\Delta S$ is produced or energy $\Delta U$ is dissipated in each infinitesimal equilibration. In the ideal 
reversible limit, these losses become zero. 
A measure of the amount of dissipation along a process is given by the notion of thermodynamic length of a process.

\section{Finite-time thermodynamics}
The thermodynamic length of a process can be defined in the framework of geometric thermodynamics, using the definition of a 
metric in the \emph{equilibrium} phase space spanned by the state variables describing the system. 
We will here use the thermodynamic metric of Ruppeiner \cite{rupp1979}, which is defined as the Hessian of the entropy of the system, 
\begin{equation} \label{eq:metricS}
  ds_{S} = \mb{g}_{S} \, dq^a \otimes dq^b = \frac{\partial^2 S}{\partial q^a \partial q^b} \, dq^a \otimes dq^b \,,
\end{equation}
with respect to its extensive thermodynamic parameters $\mb{f}=\{ q^a \}$. \\
The length of a path with respect to the entropy metric \eqref{eq:metricS} in the equilibrium space of the system is then 
defined as 
  \begin{equation}
    L_{S} = \int_a^b ds_{S} = \int_a^b \sqrt{ d\mb{f}^{\top} \mb{g} \, d\mb{f} } \,.
  \end{equation}
or, when parametrized by a parameter $\xi$, 
  \begin{equation}
    L_{S} = \int_0^{\xi_{max}} \sqrt{ \dot{\mb{f}}^{\top} \mb{g} \, \dot{\mb{f}} \, } \, d\xi \,.
  \end{equation}
The entropy production along a path between the points $a$ and $b$ can be derived as \cite{SalamonBerryPRL}
  \begin{equation}
    \Delta S = \frac{1}{2} \int_0^{\xi_{max}} \, \epsilon_S  \, \dot{\mb{f}}^{\top} \mb{g} \, 
      \dot{\mb{f}} \, d\xi =: \bar{\epsilon}_S \, \mathcal J_S \,,
  \end{equation}
where $\mathcal J_S$ is the generalized action of the path, and $\epsilon_S$ ($\bar{\epsilon}_S$) is a measure of the 
infinitesimal entropy produced in each step (averaged over all steps). By comparison with the definition of the thermodynamic 
length, and considering the Cauchy-Schwarz inequality, a bound for the entropy generated along the path is found as 
  \begin{equation}
    \mathcal J_{S} \geq \frac{L_{S}^2(\xi_{max})}{2 \xi_{max}} \,.
  \end{equation}
From this bound, we see that for $\xi_{max} \rightarrow \infty$, the generalized action and thus the dissipated entropy go to zero, 
$\mathcal J_S \rightarrow 0$. Further, the shortest thermodynamic length between two points $a$ and $b$ can be achieved by following 
the geodesics of $\mb{g}$, 
  \begin{equation}
    \mathcal J_{S} \geq \frac{L_{S}^2}{2 \xi_{max}} \geq \frac{L_{S,\,geod.}^2}{2 \xi_{max}} \,.
  \end{equation}
Ultimately, the minimum entropy for a fixed path between two points is obtained by following the path with constant speed. 

\section{Energy extraction from a Kerr black hole}
The entropy $S$ of a Kerr black hole is related to its mass $M$ and its angular momentum $J$ by the relation
\begin{equation}\label{Kentropy}
  S(M,J) = 2M^2\left(1+\sqrt{1-\frac{J^{2}}{M^{4}}}\right)\,,
\end{equation}
and it obeys the first law of black hole thermodynamics, 
\begin{equation}
  dS = \frac{1}{T} dM - \frac{\Omega}{T} dJ \,,
\end{equation}
with the entropy as the fundamental potential, and the extensive thermodynamic parameters $\mb{f}=\{ M,J \}$. \\
The thermodynamic length is calculated as 
  \begin{equation}
    L_{S} = \int_0^{\xi_{max}} \left[ - 2 \dot{M}^2 - \frac{6 M^2 \dot{M}^2 - \dot{J}^2}{\sqrt{M^4 - J^2}} + 
    \frac{4 M^6 \dot{M}^2 - J^2 \dot{J}^2}{\sqrt{M^4 - J^2}^3} \right]^{1/2} d\xi \,.
  \end{equation}
In the following, we investigate two types of processes and calculate the respective thermodynamic lengths. \\
%
%
Keeping the angular momentum constant, we will first vary the black hole mass. We will consider processes 
starting from the extremal limit, $M_0 = \sqrt{J}$, and increasing the mass up to the boundary of the stable region 
for a Kerr black hole, $M_+ \leq M_1 = \sqrt{J/\alpha_{min}}$, where $\alpha = J/M^2$, and $\alpha_{min} = 0.6813$. 
The thermodynamic length is given by the integral 
   \begin{equation}
      L_{S} = \int_{M_0}^{M_+} \left[ 4 - \frac{4 M^6 - 12 M^2 J^2}{\sqrt{M^4 - J^2}^3} \right]^{1/2} dM \,,
   \end{equation}
which was then solved numerically. The left graph in Fig.\,1 shows the resulting length as a function of $M_+$ for 
various values of the angular momentum. \\
%
%
The second type of process considers a constant black hole mass, while decreasing its angular momentum from an upper 
value $J_+ \leq J_1 = M^2$ down to zero, $J_0 = 0$, and thus describes the classic example of a Penrose process, 
where all the rotational energy is extracted from a Kerr black hole. The thermodynamic length in this case can be 
calculated analytically as 
  \begin{equation}
      L_{S} = \int_{J_0}^{J_+} \left[ \frac{M^4}{\sqrt{M^4 - J^2}^3} \right]^{1/2} dJ = 
      \frac{J_+}{M} \, ~_2 F_1 \left[ \frac{1}{2}, \frac{3}{4}, \frac{3}{2}, \frac{J_+}{M^2} \right] \,.
  \end{equation}
The result is shown in the right graph in Fig.\,1, where the length is given as a function of $J_+$. 

\begin{figure}[h]
\begin{center}
\includegraphics[width=0.45\textwidth]{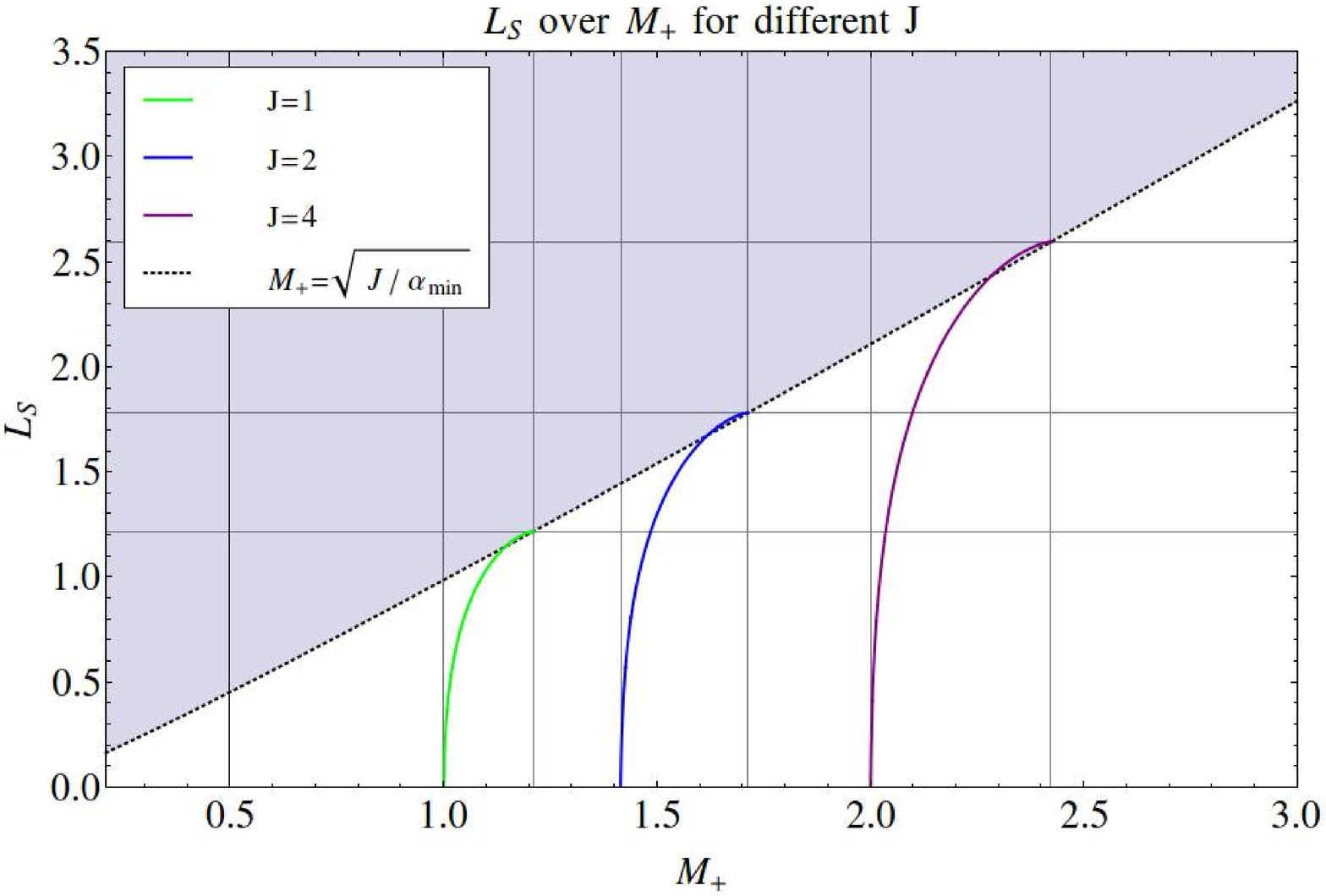} \includegraphics[width=0.45\textwidth]{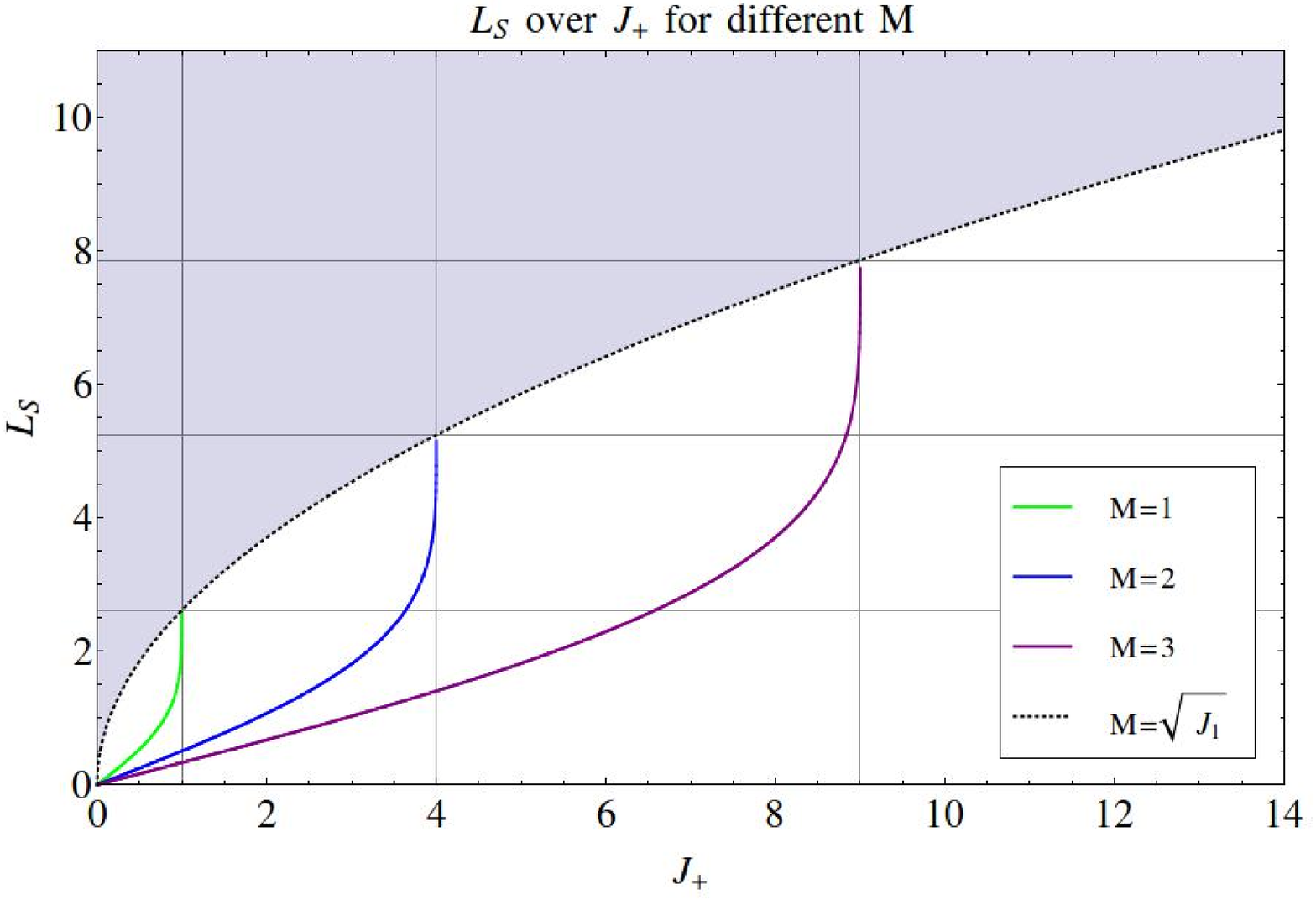}
\end{center}
\caption{(color online) Thermodynamic lengths for processes with constant $J$ (left) and $M$ (right).}
\label{aba:fig1}
\end{figure}

From the resulting curves in Fig.\,1, we deduce that close to the extremal limit the thermodynamic length grows 
stronger than further away from it, and thus the dissipation increases more strongly close to the extremal limit. For an 
efficient energy extraction therefore it is preferrable to operate away from the extremal limit, to minimize the 
dissipative losses during the process.

\section{Conclusions \& Outlook}
In this work, we considered Penrose-inspired processes, i.e. the extraction of energy fom black holes, from a thermodynamic 
point of view, taking into account the dissipative losses generated due to finite-time effects. The entropy generated along 
a process can be bound by the thermodynamic length of that process in the geometric space of equilibrium thermodynamics. 
We calculate the thermodynamic length of processes with constant angular momentum and mass of a Kerr black hole, and find 
that dissipative losses grow more strongly close to the extremal limit of the black hole. A next step is to calculate 
the length of geodesic paths in the thermodynamic geometry, minimizing the dissipative losses, which has been done in 
Ref. \refcite{2015Brav}.

\section*{Acknowledgments}
This work was done in collaboration with A. Bravetti and C. S. Lopez-Monsalvo. 
The author was supported by funding from the DFG Research Training Group 1620 `Models of Gravity'.

\bibliographystyle{ws-procs975x65}
\bibliography{GTD}

\end{document}